\documentclass[a4paper,11pt]{article}
\pdfoutput=1 

\usepackage{jinstpub} 
\usepackage{verbatim}
\usepackage{lineno}
\usepackage{subcaption}
\usepackage{tikz}
\usetikzlibrary{patterns}
\usetikzlibrary{plotmarks}
\usepackage{multirow}
\usepackage{mathtools,halloweenmath}
\usepackage{tikz}
\usepackage{overpic}
\usepackage{caption}
\usetikzlibrary{shapes.geometric, arrows}
 \tikzstyle{io} = [trapezium, trapezium left angle=70, trapezium right angle=110, minimum width=2cm, minimum height=0.8cm, text centered, draw=black]
\tikzstyle{process} = [rectangle, minimum width=2.5cm, minimum height=1cm, text centered, text width=2.5cm, draw=black]
\tikzstyle{table} = [rectangle, minimum width=4.0cm, minimum height=1cm, text centered, text width=4.0cm, draw=black]

\tikzstyle{startstop} = [rectangle, rounded corners, minimum width=2.5cm, minimum height=1cm,text centered, text width=2.5cm, draw=black]
\tikzstyle{arrow} = [thick,->,>=stealth]

\title{\boldmath Expected Performance of Cosmic Muon Veto Detector}

\author[a,b,1]{Raj Shah,\note{Corresponding author.}}
\author[b]{Gobinda Majumder,}
\author[a,c]{Prashant Shukla}    

\affiliation[a]{Homi Bhabha National Institute, Mumbai-400094, India}
\affiliation[b]{Tata Institute of Fundamental Research, Mumbai-400005, India}
\affiliation[c]{Bhabha Atomic Research Centre, Mumbai-400085, India}  

\emailAdd{raj.shah@tifr.res.in}

\abstract{

The India-based Neutrino Observatory (INO) collaboration has established a miniICAL detector, at the transit campus of IICHEP, Madurai, India, which serves as a prototype detector of the larger Iron-Calorimeter detector (ICAL). The purpose of miniICAL lies in unraveling the intricate physics and engineering challenges inherent in constructing and operating a substantial ICAL-type detector.
To explore the feasibility of building a large-scale neutrino experiment at shallow depths the collaboration has embarked upon the construction of a Cosmic Muon Veto Detector (CMVD) around the miniICAL detector. The primary objective of this endeavor revolves around attaining a veto efficiency surpassing $99.99\%$, while simultaneously maintaining a false-positive rate lower than $10^{-5}$. 
  The CMVD system is based on extruded plastic scintillators (EPS) and utilizes wavelength-shifting fibers to collect scintillation photons and uses silicon photomultipliers (SiPMs) as photo-transducers.
  A software tool is developed for CMVD and is integrated with the existing miniICAL consisting of RPC detectors. The simulation is tuned to include properties of EPSs and WLS fibers,
 measured efficiencies, and time resolutions of EPSs. Measured spectra and noise in SiPMs are also taken into account. 
 The muon tracks in the RPCs are used to estimate the muon veto efficiency of
 CMVD to arrive at efficient muon veto criteria. With improved veto efficiency of cosmic muons, the CMVD experiment will help to pave the way for future large-scale shallow-depth neutrino experiments e.g. INO-type experiments, enhancing our understanding of neutrino properties.}

\arxivnumber{2403.06114} 

\collaboration[c]{on behalf of INO collaboration}

\begin{document}
	\maketitle

	\section{Introduction}
	\label{introduction} 
	The proposed 51 kton magnetized Iron Calorimeter (ICAL) at the India-based Neutrino Observatory (INO) has the objective of accurately measuring the parameters of atmospheric neutrinos and studying the effect of matter on their oscillations. The underground laboratory, along with ICAL, is planned to be located in Bodi West hills, Theni, India, and is equipped with a rock cover of over $1\,km$ to reduce the background from cosmic-ray muons.
A rock cover of $1.3\,km$ in all directions reduces the cosmic muon flux by a factor of about $10^{6}$. Placing the detector at a depth of approximately $100\,m$ reduces the muon flux by a factor of $10^{2}$.
To achieve a reduction factor of about $10^{6}$, an active Cosmic Muon Veto Detector (CMVD) system with an efficiency greater than $99.99\%$ must be installed around the shallow-depth detector (at $100\,m$ depth).

\begin{figure}[h]
		\begin{minipage}{0.45\linewidth}
			\centering
			\includegraphics[width=\linewidth]{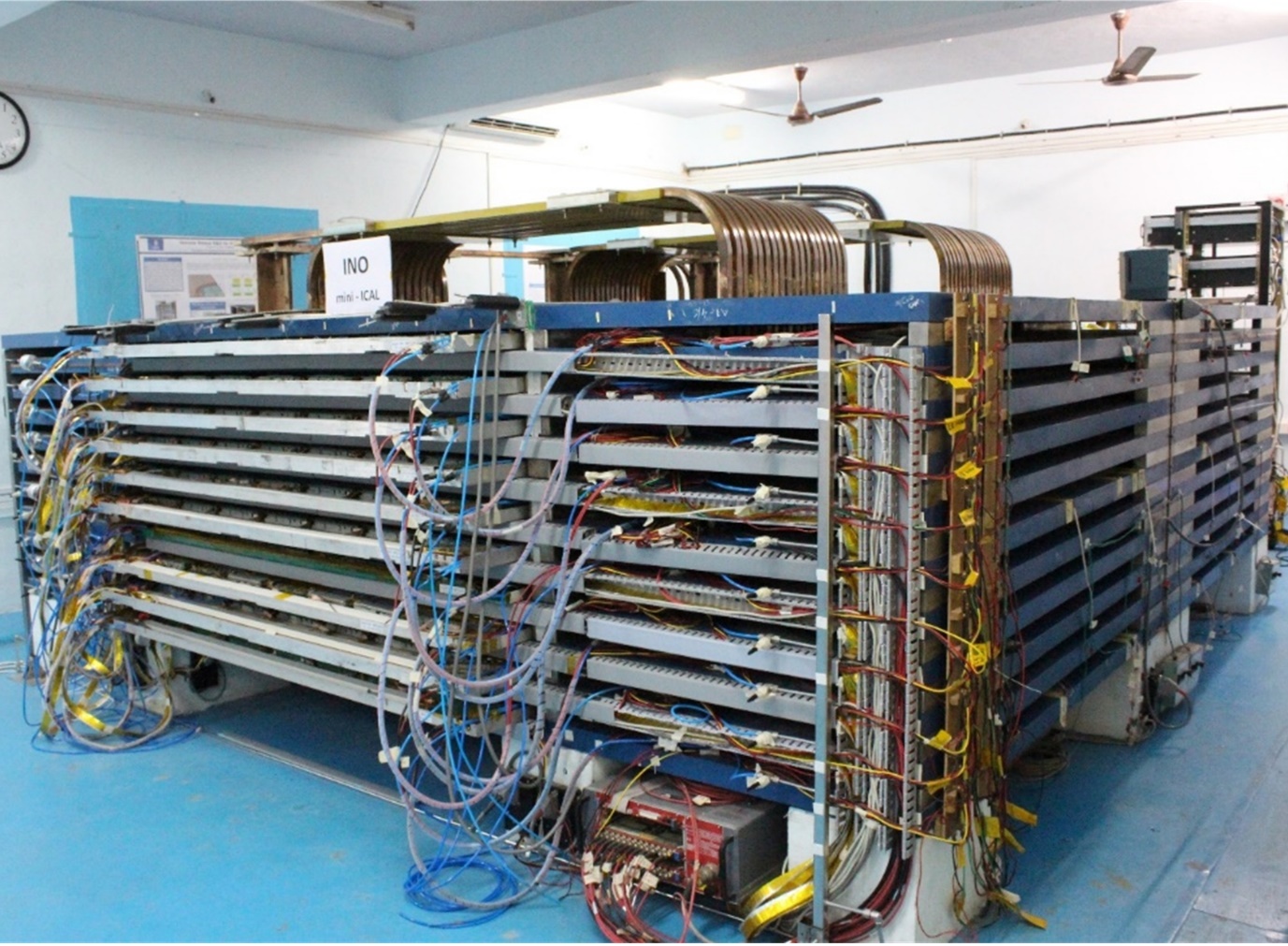}
			\caption{Fully operational miniICAL detector at IICHEP.}
			\label{mical}
		\end{minipage}
		\hfill
		\begin{minipage}{0.44\linewidth}
			\centering
			\includegraphics[width=\linewidth]{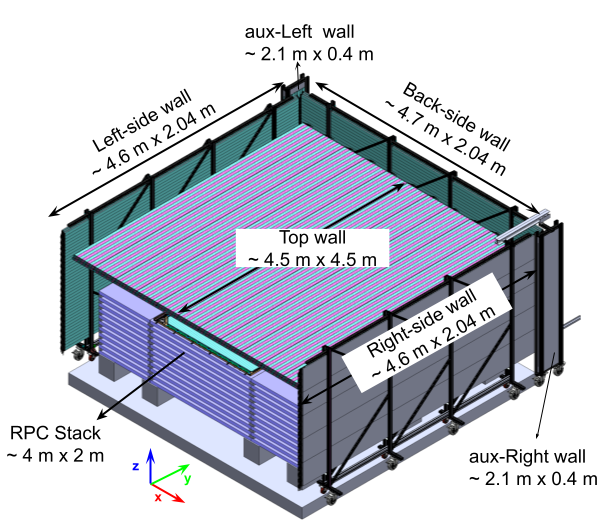}
			\caption{A sketch showing the Cosmic muon veto detector around the miniICAL detector.}
			\label{miniicalcmvd} 
		\end{minipage}
	\end{figure}

The miniICAL is a prototype of the ICAL which is operating at IICHEP Madurai. The miniICAL is a magnetized 85-ton detector consisting of 10 RPC layers between eleven, $4\,m\times 4\,m\times 5.6\,cm $ iron plates \cite{Majumder:2018xel} and a view of that system is shown
in figure \ref{mical}.
To study the performance of shallow depth neutrino detectors, the plan is to build
a cosmic muon veto detector (CMVD) on top of the existing miniICAL detector as shown in figure \ref{miniicalcmvd}. The upgrade is motivated by a previous successful demonstration of 99.98\% muon veto efficiency using a $1\,m\times\,1\,m\times0.3\,m$ veto system made from scintillator detectors with PMTs as photo-sensors \cite{Panchal:2017aub}.
	
        The cosmic muon veto detector (CMVD) around miniICAL is built using extruded plastic scintillator (EPS) strips. Each EPS strip is equipped with two wavelength-shifting (WLS) fibers for collecting scintillation photons and four silicon photomultipliers (SiPMs) as phototransducers. The performance of the EPS strips, WLS fibers, and SiPM readout systems has been thoroughly validated in prior studies \cite{Saraf:2023sjs}, \cite{Jangra:2021key}.
        
   We have developed a software tool for CMVD which simulates the muon signal in the CMVD walls and integrates with the existing miniICAL consisting of RPC detectors. The simulation is tuned to include properties of EPSs and WLS fibers, measured efficiencies and time resolutions of EPSs. The measured spectra and noise in SiPMs are also taken into account. The muon tracks in the RPCs are used to estimate the muon veto efficiency of CMVD to arrive at efficient muon veto criteria. 
	
	The paper is organized into sections, each dedicated to a different aspect of the CMVD design and performance.
	In Section 2, the geometry of the CMVD is outlined, including dimensions, composition, and arrangement.
	Section 3 gives the details the generation of cosmic muon events and the trigger criteria used to identify them.
	Section 4 focuses on the digitization of signals from SiPM which are used to detect scintillation light propagated through the WLS fiber. This Section also explains how the SiPM signal is modeled using the position and energy deposited inside the EPS.
	The track reconstruction and extrapolation techniques used to determine the path of cosmic muons through the CMVD are described in Section 6. The algorithms to reconstruct muon tracks and extrapolation to the CMVD are discussed in Section 7.
	The expected performance of the veto detector is presented in Section 9 and then conclusion are given in Section 10.

\section{Detector Geometry}
 \label{chap_geometry}
\begin{figure}[h]
	\includegraphics[width=1.\textwidth]{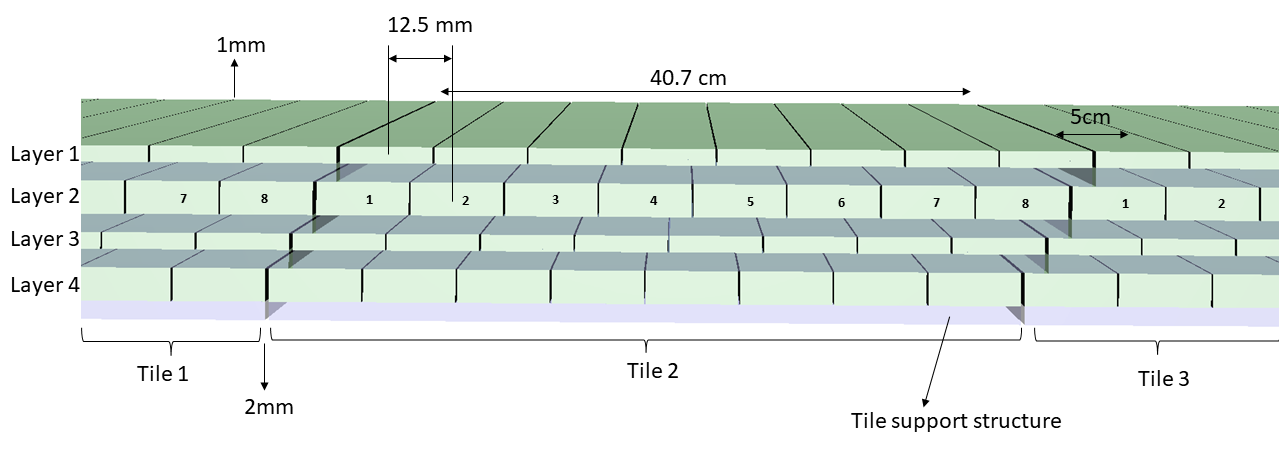}	
	\caption{Assembly of top veto wall.}
	\label{topwallassembly} 	
\end{figure}

The cosmic muon veto detector around miniICAL consists of four large-sized veto walls (Top, Left, Right, and Back) and two auxiliary walls (aux-Left and aux-Right). It utilises extruded plastic scintillator strips (EPS) \cite{Pla-Dalmau:2000puk} with dimensions optimized for maximum coverage. The scintillation light produced by incoming particles is collected through Kuraray Y-11 wavelength shifting (WLS) fibers\cite{Alekseev:2021vbe} embedded in the EPSs and read out on both sides using $2\,mm \times 2\,mm$ Hamamatsu Silicon Photomultiplier (SiPM) S13360-2050VE \cite{sipm}.	\\
The existing miniICAL geometry in GEANT4 toolkit \cite{Agostinelli:2002hh} has been modified to incorporate the cosmic muon veto detector.
Each wall is constructed using up to 4 layers of EPSs, which are designed in the form of tiles as shown in Figure \ref{topwallassembly} for the top wall.
Each of these layers are arranged in a staggered configuration to avoid the loss of efficiency due to dead spaces between strips. 
Each individual tile is created by placing 8 EPS strips side by side. In order to streamline electronic channels and improve efficiency, two of these strips are combined into a single unit referred to as a "di-counter" \cite{Saraf:2023sjs}.
The gap between each layer is $10\,mm$ for the mechanical support structure (aluminium tile base), while the gap between EPS strips is $\sim1\,mm$ and that between two tiles is $\sim2\,mm$, due to glueing, coating, mechanical tolerances and packing materials. 
 
%
%

\begin{figure}[htb]
	\centering
	\begin{subfigure}[b]{\textwidth}
		\centering
		\includegraphics[width=0.44\textwidth]{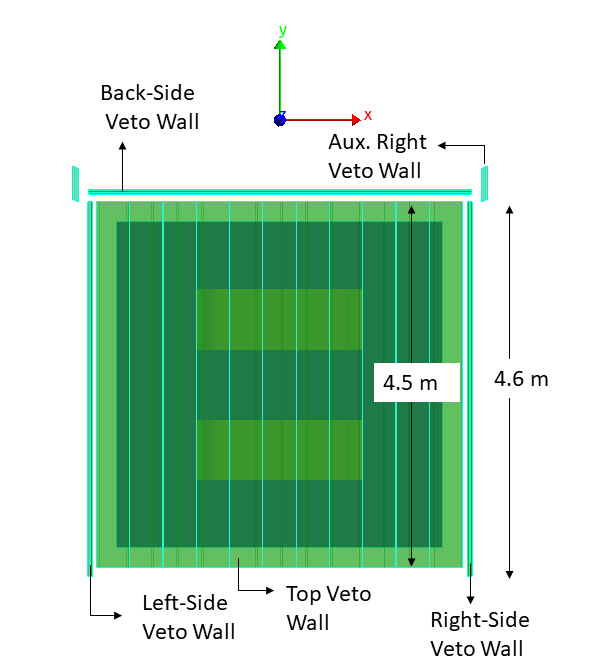}
		\includegraphics[width=0.48\textwidth]{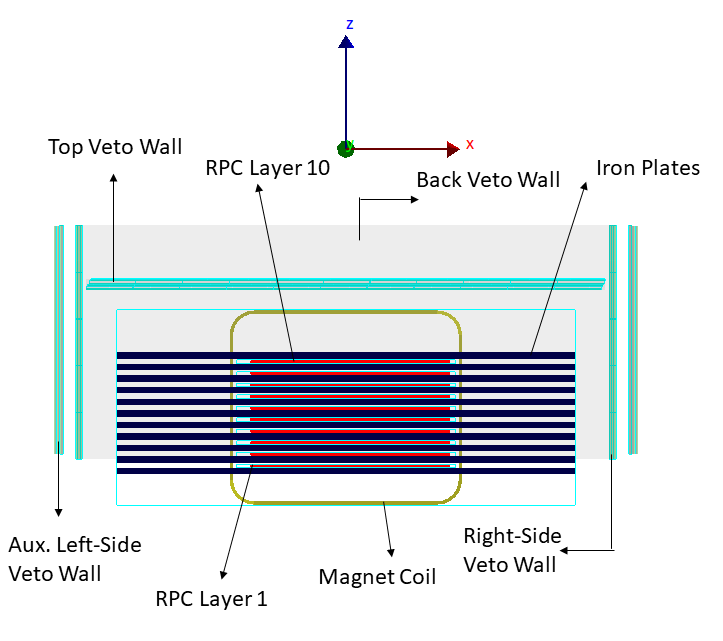}
	\end{subfigure}
	
	\caption{(left) Top and (right) front view of the cosmic muon veto detector showing top and side veto walls.}
	\label{fig:geant4cmvd}   
	
\end{figure}
Figure \ref{fig:geant4cmvd} shows top and front views of simulated CMVD geometry around miniICAL in GEANT4.
 The top-veto wall is positioned at $52\,cm$ from the topmost iron layer of miniICAL and is equipped with EPS strips, which run along the y-axis for length and x-axis for width. The left and right-side walls are positioned at roughly $26\,cm$ from the miniICAL detector, with EPS strips running along the y-axis for length and z-axis for width. The back wall is located at $33\,cm$ from the miniICAL detector, the EPS strips are running along the x-axis for length and z-axis for width. Additionally, two small-sized veto walls are placed near the edges between the back and left/right walls to veto muons passing through the gap between edges of back and side-walls. The positions of all veto walls have been optimised so that when miniICAL is triggered by a muon trajectory, it passes through at least one veto wall, except in the front side.  The side veto walls are elevated $25\,cm$ from the floor for better coverage of miniICAL from top and sides.
 Table \ref{table:cmvdspec} summarizes the general parameters of EPS used in the veto detector \cite{Saraf:2023sjs}. In total there are 748 such EPS strips making up the veto detector. In simulation, these EPS strips are made up of polystyrene with a density of 1.05\,$g/cm^{3}$. 

\begin{table}[htbp]
\begin{tabular}{|c|c|c|c|c|c|}
\hline
Veto  & $\#$ of   & Layer  & $\#$ of  & Strip & distance \\
Wall & Layers  &staggering &Tiles & dimensions & from miniICAL \\ \hline
Top & 4  &  $\frac{1}{4} $ width& 11/Layer & $4.5\,m\times 5\,cm \times 1$ - $2\,cm$ & $ 52\,cm $ \\ 
\hline
Side & 3 & $\frac{1}{3} $ width& 5/Layer & $4.6\,m\times 5\,cm \times 1\,cm$  & $ 26\,cm$\\  \hline
Back & 3 & $\frac{1}{3} $ width& 5/Layer & $4.7\,m\times 5\,cm \times 1\,cm$  & $ 30\,cm$\\  \hline 
Auxiliary & 3 &  $\frac{1}{3} $ width& 1/Layer & $2.1\,m\times 5\,cm \times 2\,cm$ & $ 33\,cm $ \\ \hline
\end{tabular}
\caption{Geometrical specifications of cosmic muon veto detectors \cite{Saraf:2023sjs}.}
\label{table:cmvdspec}
\end{table}
\section{Event Generation and Simulation Steps} 
\label{event_generation} 


The Monte-Carlo event generation was performed using extensive air shower simulation framework CORSIKA \cite{Heck:1998vt} while the simulation of the passage of a particle through the detector geometry is performed by the GEANT4 simulation toolkit. 
The trigger criteria used to collect cosmic muon data in miniICAL is that muon must have signals in the top four RPC layers \cite{John:2022fuy}. The same criteria is used in the simulation.
Event generation process starts with a random position ($x$, $y$) in the sensitive area of the top RPC layer (layer 10) and momentum component ($P_{x}$, $P_{y}$ and $P_{z}$) extracted randomly from the output of CORSIKA. 
Then, the particle's position is projected downwards to the bottom RPC layer (layer 7) to verify that the muon has passed through all these four layers to satisfy trigger criteria.
Lastly, the particle's trajectory was projected upwards to the roof to determine its starting point, referred to as the vertex.

\begin{figure}[h]
	\centering
	\includegraphics[width=0.5\linewidth]{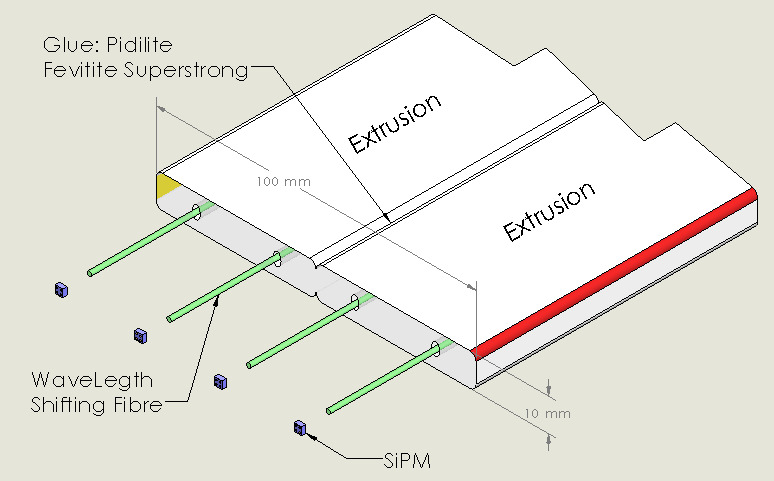}
	\caption{Exploded view of the di-counter's one end showing four SiPMs aligned with corresponding WLS fibers.} 
	\label{diCounter_Assembly}
\end{figure}

When a charged particle passes through the EPS strip, it produces scintillation light which is then absorbed, re-emitted inside the wave-length shifting fibers, and propagated to the SiPMs. To maximize light collection, each EPS strip has two fibers, resulting in four SiPMs per EPS. Figure \ref{diCounter_Assembly} shows one side-view of a di-counter with four SiPMs belonging to two different EPS. The other four SiPMs are mounted on the other side of the di-counter. Fibers are extended beyond the EPS end to accommodate the size of the readout electronics boards of the SiPMs. 
At the strip level, one has information about the time, position, and energy deposited when a muon passes through the EPS strip while at the SiPM level, one has digitised information about the timing and the integrated charge in all 4 SiPMs associated with the particular EPS. The signal in each SiPM depends on the energy deposited and the distance between the muon hit and the end of the WLS fiber, where the SiPM is mounted.
  The digitized charge and timing information from Geant4 output
  is fine-tuned to replicate the SiPM signal obtained from standalone cosmic muon setup
with EPS as described in Ref.~\cite{Saraf:2023sjs}.

 \subsection{Charge Measurement}

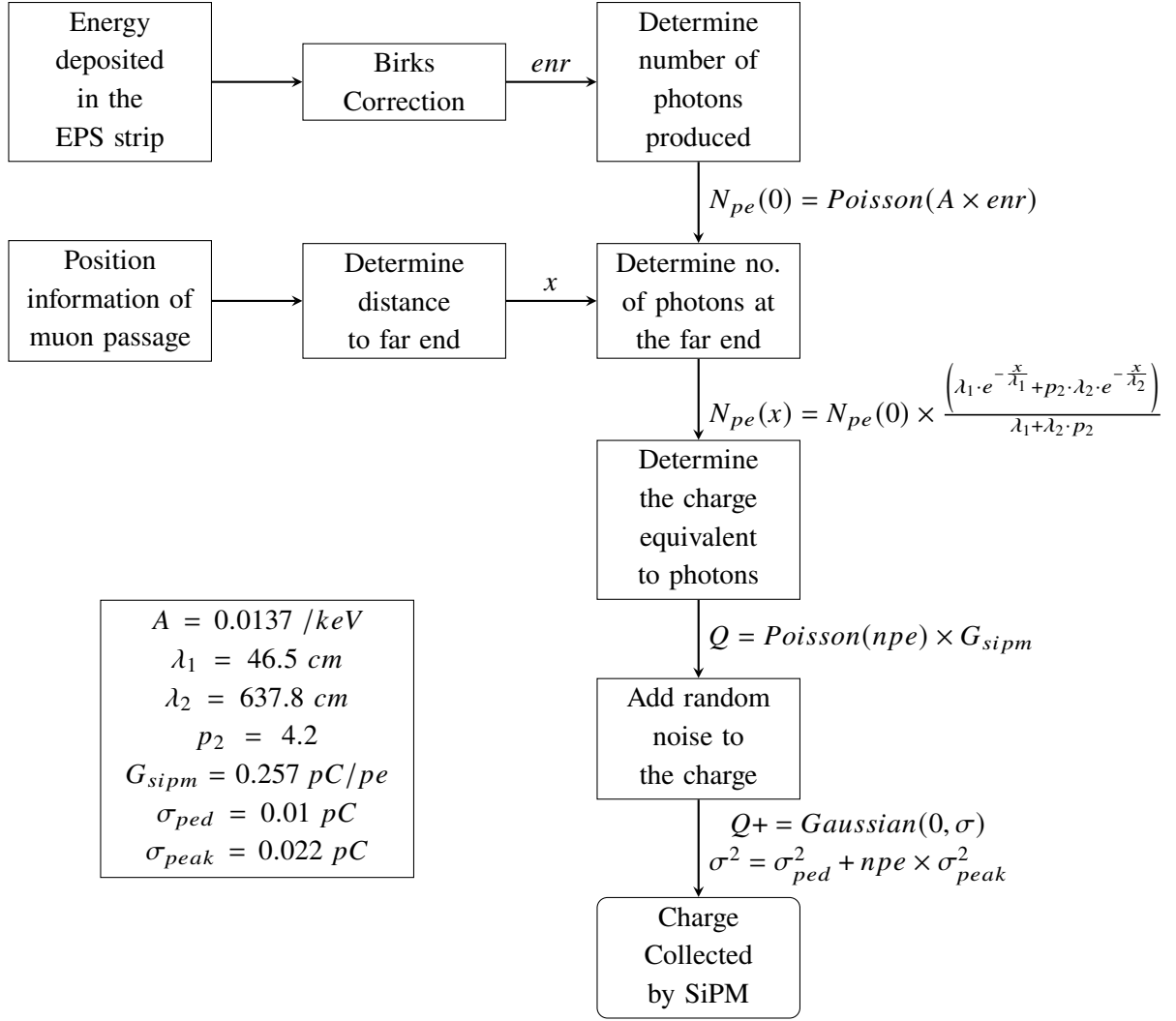
\begin{figure}[h]
\centering
\begin{tikzpicture}[node distance=1.cm]
\node (pro2) [process] {Determine\\ number of\\ photons produced};
\node (pro1) [process, left of=pro2, xshift=-3.cm] {Birks \\ Correction};
\node (input1) [process, left of=pro1, xshift=-3.cm] {Energy deposited \\ in the EPS strip};
\node (pro4) [process, below of=pro2, yshift=-2.cm] {Determine no. of photons at the far end};
\node (pro3) [process, left of=pro4, xshift=-3.cm] {Determine distance to far end};
\node (input3) [process, left of=pro3,xshift=-3.cm] {Position information of muon passage};
\node (pro5) [process, below of=pro4, yshift=-2.cm] {Determine the charge equivalent to photons};
\node (pro6) [process, below of=pro5, yshift=-2.cm] {Add random noise to the charge};
\node (startstop)[startstop, below of=pro6,yshift=-2.cm] {Charge\\ Collected\\ by SiPM};
\node (parameters)[table, right of=pro6,xshift=-7cm] {$A = 0.0137 \:/keV $ \\ $ \lambda_{1} = 46.5\:cm$ \\ $ \lambda_{2} = 637.8\:cm $ \\ $p_{2} = 4.2$  \\ $G_{sipm} = 0.257 \:pC/pe $ \\ $\sigma_{ped}= 0.01\:pC$ \\ $\sigma_{peak}= 0.022\:pC$};

\draw [arrow] (input1) -- (pro1);
\draw [arrow] (pro1) --node[anchor=south] {$enr$} (pro2);
\draw [arrow] (pro2) -- node[anchor=west] {$N_{pe}(0) = Poisson (A \times enr)$} (pro4);
\draw [arrow] (pro3) -- node[anchor=south] {$x$} (pro4);
\draw [arrow] (input3) -- (pro3);
\draw [arrow] (pro4) --node[anchor=west] {$N_{pe}(x) = N_{pe}(0) \times \frac{ \left( \lambda_1 \cdot e^{-\frac{x}{\lambda_1}} + p_2 \cdot \lambda_2 \cdot e^{-\frac{x}{\lambda_2}} \right)}{\lambda_1 + \lambda_2 \cdot p_2}$} (pro5);
\draw [arrow] (pro5) -- node[anchor=west, align=center] {$Q = Poisson(npe) \times G_{sipm}$} (pro6);
\draw [arrow] (pro6) -- node[anchor=west, align=center] {$Q+=  Gaussian(0, \sigma)$ \\ $\sigma^2 = \sigma_{ped}^2 + npe \times \sigma_{peak}^2$} (startstop);
\end{tikzpicture}
 \caption{Flowchart of the Monte Carlo simulation for charge measurement in SiPM.}
  \label{fig:algorithm_flowchart}
\end{figure}

 A Monte Carlo-based model was developed to simulate the charge response of a SiPM based on the energy deposition by muons in an EPS and their positions.
  Figure \ref{fig:algorithm_flowchart} shows the overall algorithm of calculating the charge of the SiPM using the energy deposited and the position information of muon hit inside the EPS.
 To account for the saturation in the plastic scintillation light yield, the energy deposited per unit length $\frac{dE}{dx}$ is scaled using Birk's formula \cite{Birks:1951boa} ($kB = 0.126\,mm/MeV$ for polystyrene).
  The gain $G_{sipm}$ of the SiPM, pedestal width, $\sigma_{ped}$, and the width of single photoelectron peak, $\sigma_{peak}$, were obtained by characterizing the SiPMs using LED source \cite{Jangra:2021key}.  


  The photon intensity attenuation inside EPS and WLS fibers have been measured using a test setup
described below.
Two small EPS paddles were placed orthogonally to three long EPSs on both sides.
The two paddles and two of long EPS detectors were used to trigger the muon signal to test the third long EPS. By varying the position of the smaller EPSs, the muon's position from the signal end was varied, within the uncertainty of the overlapped area.

\begin{figure}[h]
  	\centering
        \includegraphics[width=0.65\linewidth]{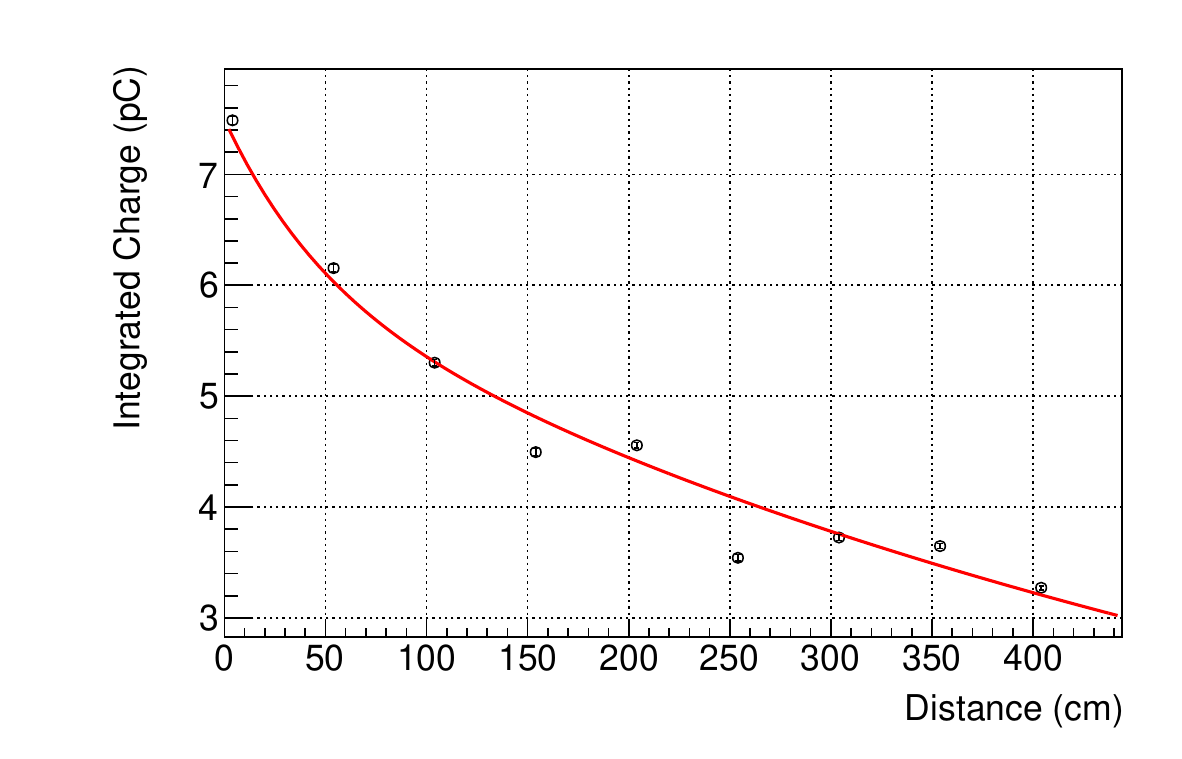}
  	\caption{Integrated charge in a SiPM as a function of the distance of the muon position inside the EPS.} 
  	\label{atten}
  \end{figure}

Integrated charges in the SiPM for all these data were fitted with a Gaussian convoluted Landau function to extract the peak of the Landau function.
The peaks at different positions are then fitted with a double exponential function $ A [exp(-\frac{x}{\lambda_{1}}) + f \times exp(-\frac{x}{\lambda_{2}})]$, as depicted in Figure \ref{atten}, where noisy data at $150$ and $250\,cm$ were excluded from the fit. The fit parameters are $ A = 1.44 \pm 0.04\,pC $, $ \lambda_{1} = 46.5 \pm 3.8\,cm $ and $\lambda_{2} = 637.8 \pm 8.5\,cm $, with the fraction, $f$ fixed at 4.2.
The resulting average number of photoelectrons (p.e.) generated at the far end of the EPS is determined to
 be $0.013\,p.e./keV$ which is used in the simulation.
This model was implemented in the simulation but was verified with the di-counter testing setup \cite{Saraf:2023sjs} in GEANT4.
Total 20k events with similar trigger conditions as described in Ref.~\cite{Saraf:2023sjs} were simulated.
    
\begin{figure}[h]
    \centering
    \begin{subfigure}{0.45\textwidth}
        \centering
        \begin{overpic}[width=\linewidth]{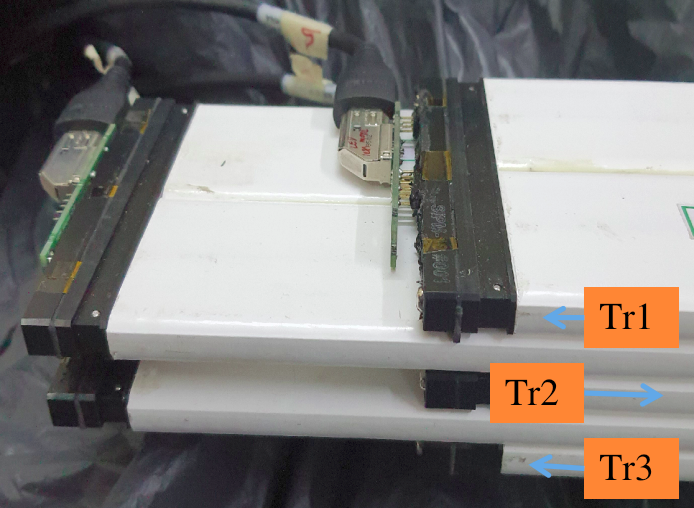}
	       	\put(50, 55){\parbox{0.9\linewidth}{\captionsetup{font=Large}\caption{}}}
        \end{overpic}
        \label{fig:experimental_setup}
    \end{subfigure}
    \hfill
    \begin{subfigure}{0.45\textwidth}
        \centering
        \begin{overpic}[width=\linewidth]{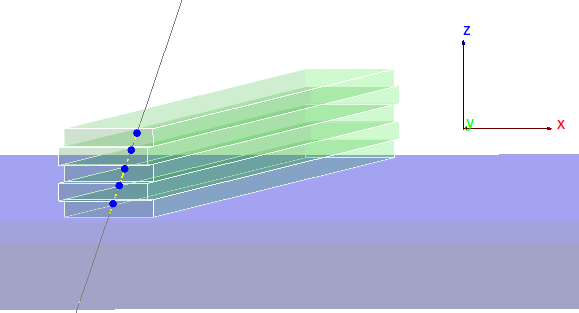}
        	\put(50, 50){\parbox{0.9\linewidth}{\captionsetup{font=Large}\caption{}}}
        \end{overpic}
    
        \label{fig:simulation_geometry}
    \end{subfigure}
    \caption{Di-counter test setup: (a) Photograph of the experimental setup. Tr1, Tr2, and Tr3 are used to generate cosmic muon trigger \cite{Saraf:2023sjs},  (b) Diagram of the GEANT4 simulation geometry.}
    \label{fig:type3}
\end{figure}

\begin{figure}[h]
  	\begin{center}
  		\includegraphics[width=0.7\textwidth]{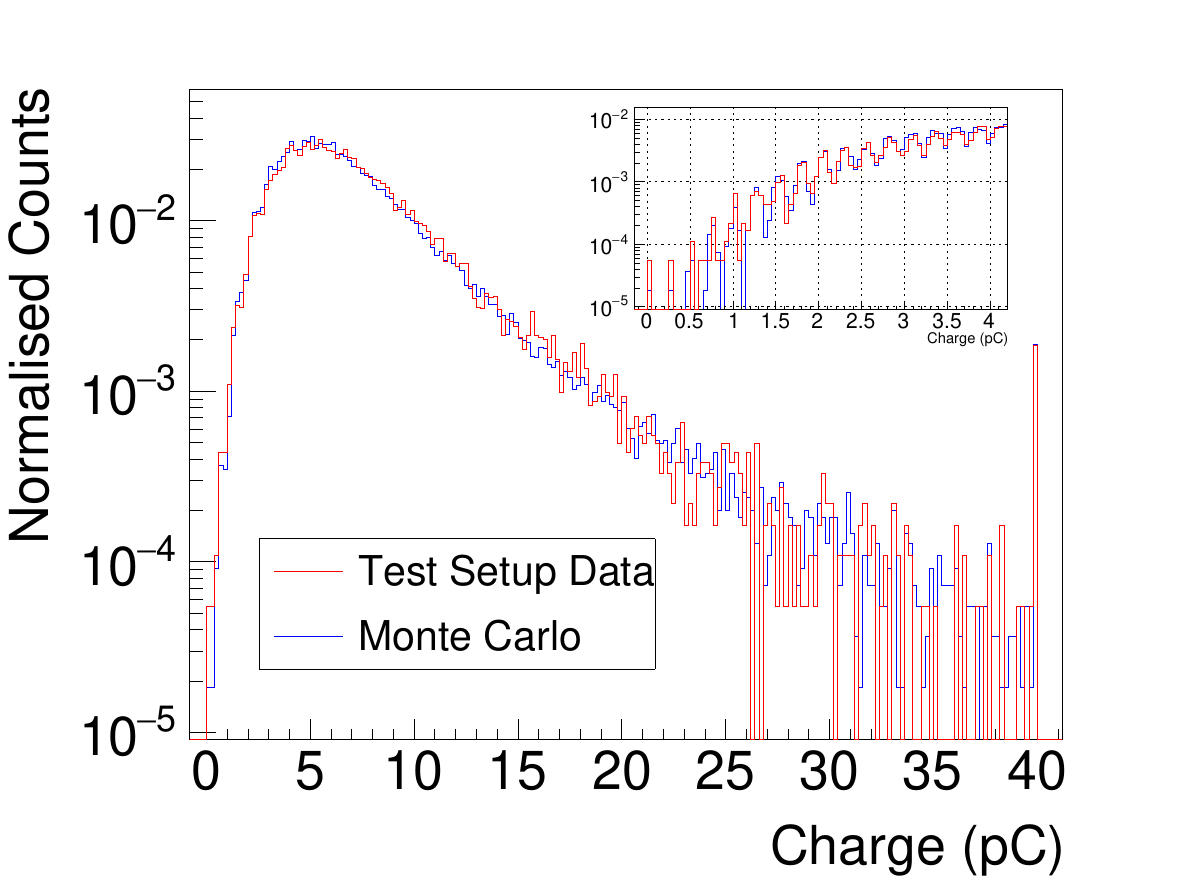}
  		\caption{Simulated (blue) and real charge spectrum (red) of one of the SiPMs in di-counter testing setup. Insert is the zoomed-in view near the pedestal.}
  		\label{fig:chargespectrum}
  	\end{center}
  \end{figure}

Figure \ref{fig:type3} (a) shows the di-counter test setup where two test di-counters ($4.7\,m$) are inserted between three trigger di-counters of slightly smaller length ($4.5\,m$) to trigger muon events anywhere along $4.5\,m$ length in order to study the muon signals along the EPS. 
Only the front-side EPS in each trigger di-counter's is used for generating cosmic muon triggers.
Figure \ref{fig:type3} (b) shows the Di-counter geometry as implemented in GEANT4 simulation.
Figure \ref{fig:chargespectrum} compares the observed and simulated charge responses of one of the SiPM in the di-counter setup for the whole length of $4.4\,m$, where the MC prediction matches well with data.

  
\subsection{Timing Measurement}
  
The observed SiPM time signal is a convolution of photon propagation along the scintillator strip \cite{Jangra:2021key} and of the uncertainty on the measurement of time in the SiPM device \cite{Saraf:2023sjs}.

\begin{figure}[h]
\centering
\includegraphics[width=0.9\textwidth]{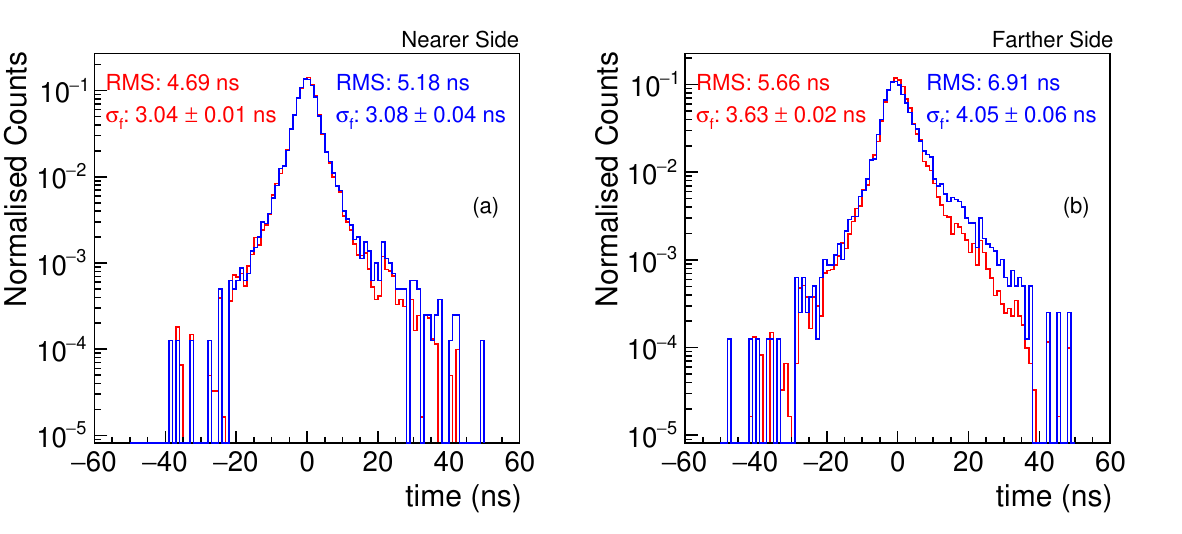}
 \caption{Timing resolution comparison between test-setup data (blue) and simulation (red) for nearer (a) and farther (b) side channel.}
 \label{fig:timereso} 
\end{figure}

The timestamp is obtained from the coincidence of four EPS signals.
  The uncertainty on the timing mainly depends on the time constant of EPS and WLS, propagation of light in the WLS fiber, and on the characteristics of SiPM and amplifier. With the increase in the number of pe/large signals, the time resolution is improved.
  Figure \ref{fig:timereso} shows the comparison of timing information in data and Monte Carlo after mean correction for both the nearer and farther SiPM,
with the closer SiPM approximately $1\;m$ away and the farther one about $3.5\:m$ away.

  The time resolution obtained is $4.00\pm 0.06\:ns$ for farther channels and $3.08\pm 0.04\:ns$ for the closer channels.
The corresponding position uncertainty from the core of the timing resolution is $36.76 \pm 1.13\:cm$ after correcting for the width of the trigger paddle ($5\:cm$) and jitter in the trigger signal ($\approx 1.5\:ns$). The tail of the timing distribution will give much larger uncertainties in the position measurements for the signal of lower integrated charge. The large discrepancies in time resolution for larger distances primarily comes from the signals of low integrated charge, which has a large variation due to the large time constant of WLS fibers.

\subsection{Integrated Noise in SiPM}
\label{chap_noise}
\begin{figure}[h]
	\centering
	 		\includegraphics[width=0.6\textwidth]{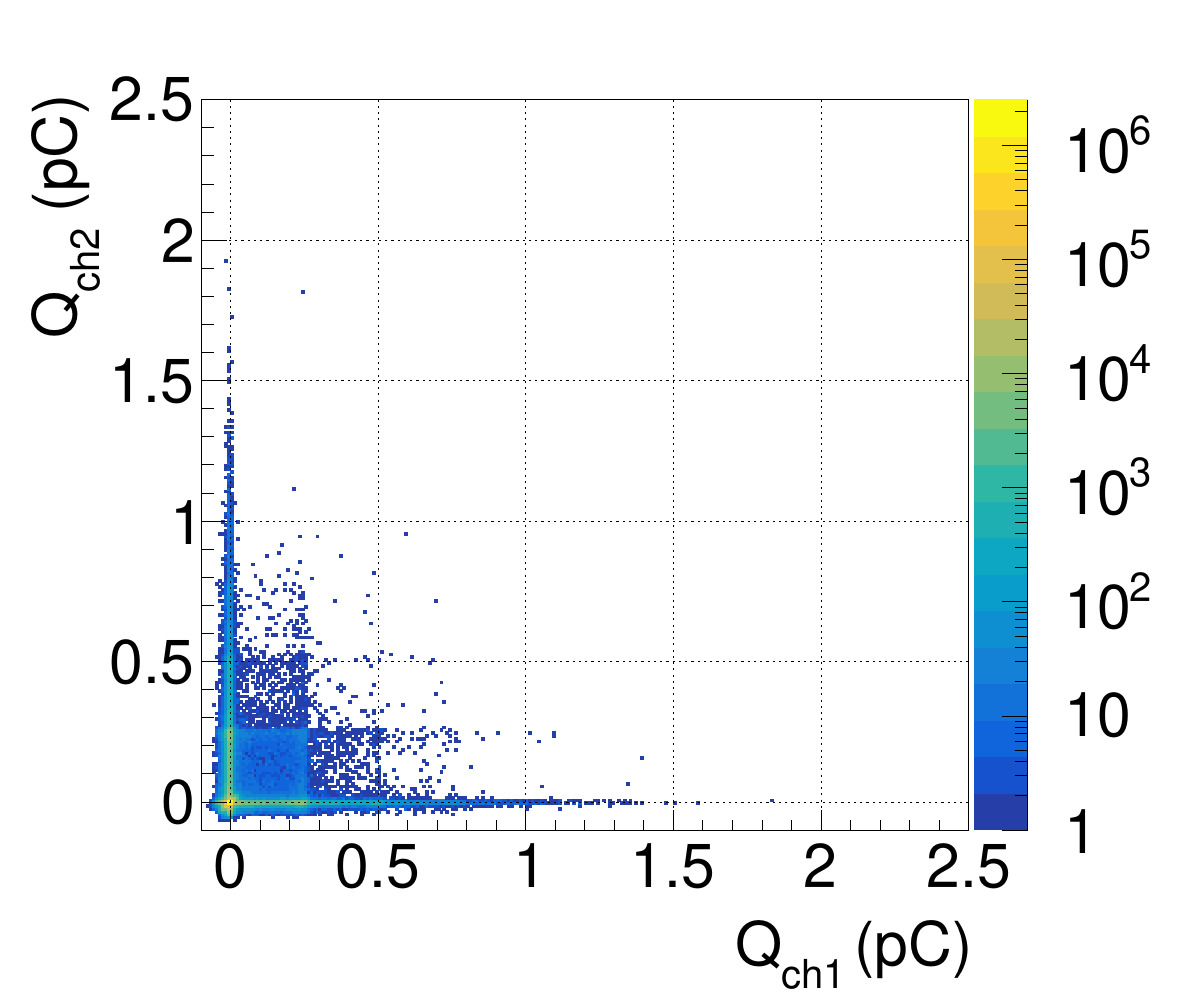}
	\caption{Correlated integrated noise between two adjacent SiPMs connected to the same electronic board.}
	\label{corr_noise}
\end{figure}

In the upcoming veto system, approximately 3000 SiPMs will be utilized. In addition to detecting muon signals, these SiPMs will also produce signals due to dark noise. Previous studies have already examined the effect of noise rates on the veto efficiency \cite{Jangra:2021key}. These SiPMs will operate at room temperature with a 2.5 overvoltage ($V_{ov}$), and it is required that at least two SiPMs register a signal above 2.5\,p.e. In the experiment, four SiPMs will be aggregated onto an adapter board for all electrical and readout connections. Subsequently, the signals will be  amplified  using  transimpedance amplifiers. For this study, the noise data was acquired in a dark environment using a DRS4 \cite{drs4}, at 2.5 $V_{ov}$ and room temperature, mirroring the conditions of the upcoming CMVD experiment. The DRS4 was triggered randomly, and the charge was integrated within randomly selected $100 \,ns$  time windows. This noise charge will be identical to what is expected in the data triggered by the miniICAL. The choice of $100 \,ns$  time window is taken as more than 90\% of muon pulse observed in the SiPM is contained within this duration. The correlated noise between two SiPMs in an adapter board is shown in figure \ref{corr_noise}. In the current simulation, integrated noise is introduced to all 3000 SiPMs from the test-setup data.

	Due to the fluctuation of all these processes and a given threshold for storing the
signal of SiPM signal, there is a possibility that SiPM(s) associated with an
EPS does not have any valid digitised signal even though a muon passed
through the EPS. Similarly, due to the fluctuation in the noise, there might
be a signal in the SiPM without any muon in the associated EPS.
At least two SiPMs associated with an EPS must have digitised signal above the certain
threshold in terms of the required p.e. to reconstruct muon position in the EPS.

In the simulation two position coordinates are obtained from the layer number and the strip number of the EPS having a valid hit while the coordinate along the length of the EPS is not considered due to larger uncertainties in the timing.



\section{Hits, Cluster and Supercluster Formation}
\label{chap_hitandcluster}

To meet the minimum efficiency criteria of $99.99\%$ and fake rate less than $10^{-5}$, two or more SiPMs in an EPS must have a signal above 2.5 p.e. \cite{Jangra:2021key}, which are defined as ``hits''. This primarily eliminates the dark noise of SiPM.  
The geometry of the veto system has been designed to detect a minimum of two layers of valid hits for every reconstructed muon in the miniICAL, excluding the front side. While a typical muon passage generates a single hit in a layer, closely spaced strips can result in two or more hits. Additionally, two or more hits can also occur due to the combination of ionization and delta rays or from muon-nuclear interactions of primary muons with detector material. Thus a cluster is formed by combining multiple adjacent strips in the same layer. The cluster position is determined by averaging the individual hit positions.



  
\begin{figure}[h]
	\includegraphics[width=1\textwidth]{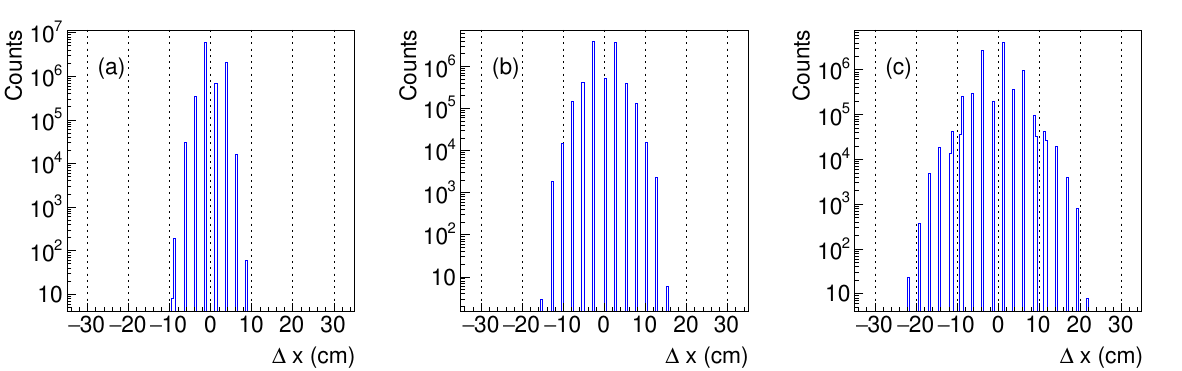}
	
	\caption{Differences in cluster $x$-positions between layer 1 and (a) layer-2, (b) layer-3 and (c) layer-4, in the top wall.}
	\label{fig:supclcut}
\end{figure}

Clusters in different layers within the same wall are combined into doublets if their separation falls within a specific distance criterion. This criterion is established by exclusively considering hits only from muons in the simulation i.e., excluding SiPM noise, and disregarding hits formed from secondary particles induced by muon interaction with detector material.  Figure \ref{fig:supclcut} illustrates the distances between clusters belonging to the 1st layer and subsequent layers in the top wall.
The width of the distributions arises primarily from the muon inclination angle and the staggering of different layers. Figure \ref{fig:supclcut} (c) exhibits a larger distance between two clusters due to the increased distance between the layers. 
Any two layers in the same wall are considered for doublet formation to account for the gap between scintillator strips and inefficiencies. The doublet position is determined as the average position of the two clusters.
\begin{figure}[h]
 	\centering
 	\includegraphics[width=0.4\textwidth]{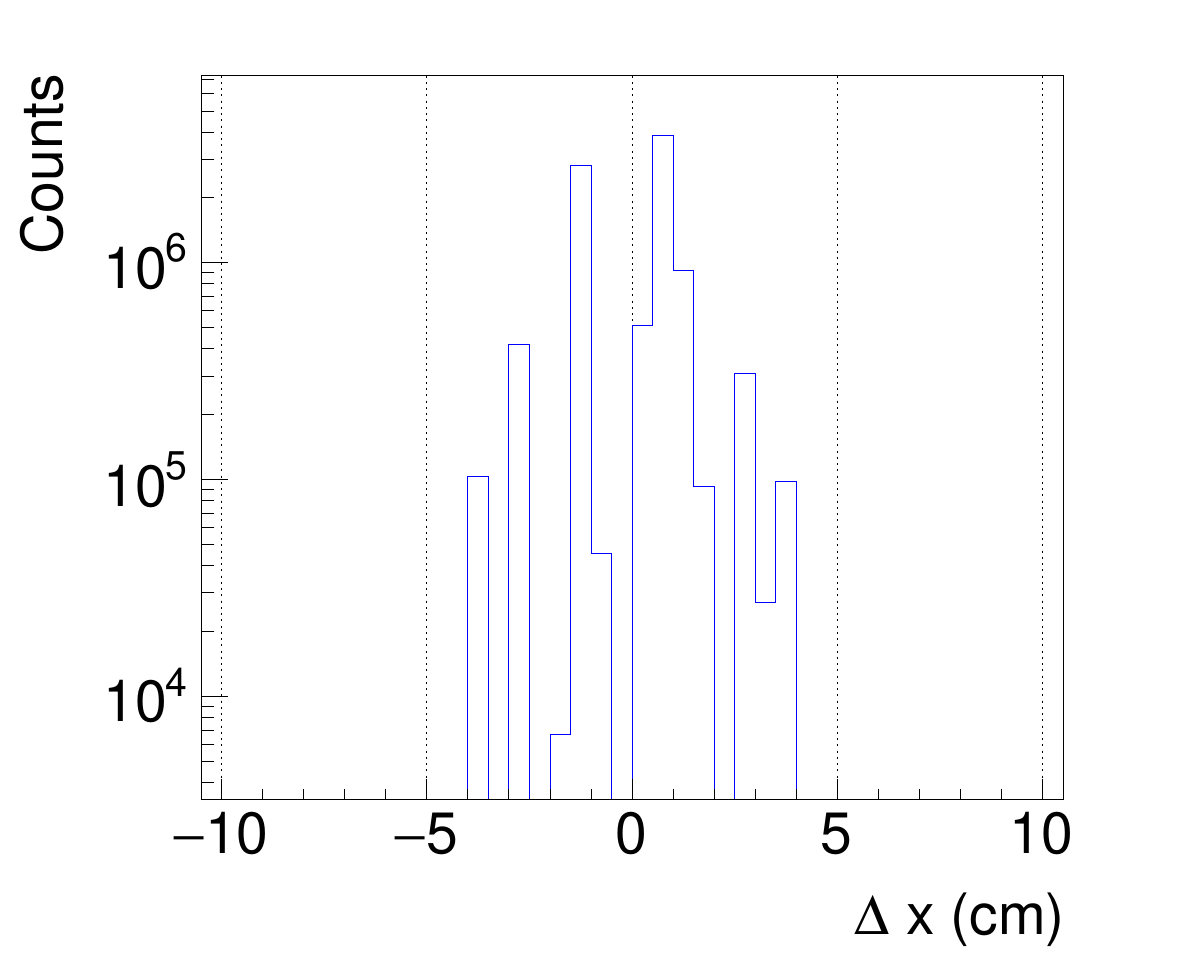}
 	
 	\caption{Differences in $x$-positions between two doublets belonging to layer-1 in the top wall.}
 	\label{fig:doubletcut}
 \end{figure}
 
  Multiple doublets are combined to form triplets if they originate from the same layer (i.e.  the initiation of one doublet and the termination of another lie on the same layer) and their difference is less than $5\,cm$ %
  as shown in figure \ref{fig:doubletcut}. For the top wall, triplets and doublets are combined to form quartets since there are 4-layers in the top wall. 
  Finally, all clusters that have been utilized in the grouping processes are discarded, and the resulting triplet/quartet is denoted as a "supercluster".
 These superclusters are fitted with straight-line and expected muon position is determined at the center (e.g., for the top wall, the center is at midpoint of the z-coordinate of the four layers). The expected muon positions are then compared with the extrapolated position derived from the muon track obtained from the RPC stack (see Section \ref*{chap_extrapolation}).
  
 



The fake rate in the efficiency measurements has to be less than $10^{-5}$ to achieve the desired veto efficiency. The SiPMs are categorized as noisy under two conditions, $(i)$ during simulation, a signal of SiPM is found to originate from noise and $(ii)$ a muon signal falls below the threshold and, after incorporating noise, the total signal surpasses this threshold. While combining the four SiPMs belonging to the same EPS, the hit is marked as noisy if three or more SiPMs are noisy.
  If all the hits comprising a cluster are noisy then the cluster is marked noisy. Moreover, if two or more clusters in a supercluster originate from noise then the supercluster is marked noisy. This identification helps to estimate the fake rate originating from intrinsic SiPM noise.

\section{Reconstruction and Extrapolation of Muon Tracks}
\label{chap_extrapolation}  
The miniICAL simulation algorithm employs a two-step process for track reconstruction, consisting of a track-finding algorithm followed by a track-fitting algorithm. The track fitting algorithm uses information about the local magnetic field to reconstruct both the momentum and charge of the muon. The Kalman filter-based track fitting algorithm is used to estimate the track parameters and its error matrix \cite{Bhattacharya:2014tha}. However, in the absence of the magnetic field, a simplified linear least-square method is implemented \cite{Pal:2014tre}. 
The track fitting algorithm provides an estimate of fitted-track parameters, $x$, $y$, $\frac{dx}{dz}$, $\frac{dy}{dz}$
  (also charge$/$momentum ratio in presence of magnetic field) at the topmost RPC plane having a valid signal.
  The next step is the extrapolation of the reconstructed track towards the veto walls, which also is done in two steps: inside the magnetic field region (inside the upper iron layers), the tracks are extrapolated using the prediction step of Kalman-filter based algorithm to propagate track parameters outside the topmost iron layer, where the magnetic field is negligible. Subsequently, beyond the iron layer, we employ a simple line-plane intersection technique.
 
  Each of the six veto planes is defined by four corners and direction cosine (normal vector to the plane). Initially, the extrapolation algorithm checks for track intersection with all these six walls, but for large zenith angles, the intersections points can be far away for vertical walls, particularly the back wall, whereas the uncertainty in the direction of muons in miniICAL is very small. Thus, to address this, the algorithm is modified
  to check the distance of the closest approaches between the extrapolated track and the line, which is formed parallel to the EPS length passed through the reconstructed supercluster position in the wall. In scenarios involving multiple superclusters within the same wall, the algorithm selects the one with the least distance of closest approach from the reconstructed muon track. Additionally, when multiple walls house valid superclusters, priority is given to the one with the least distance of the closest approach.
   \begin{figure}[h]
  	\centering
  		\includegraphics[width=1.\textwidth]{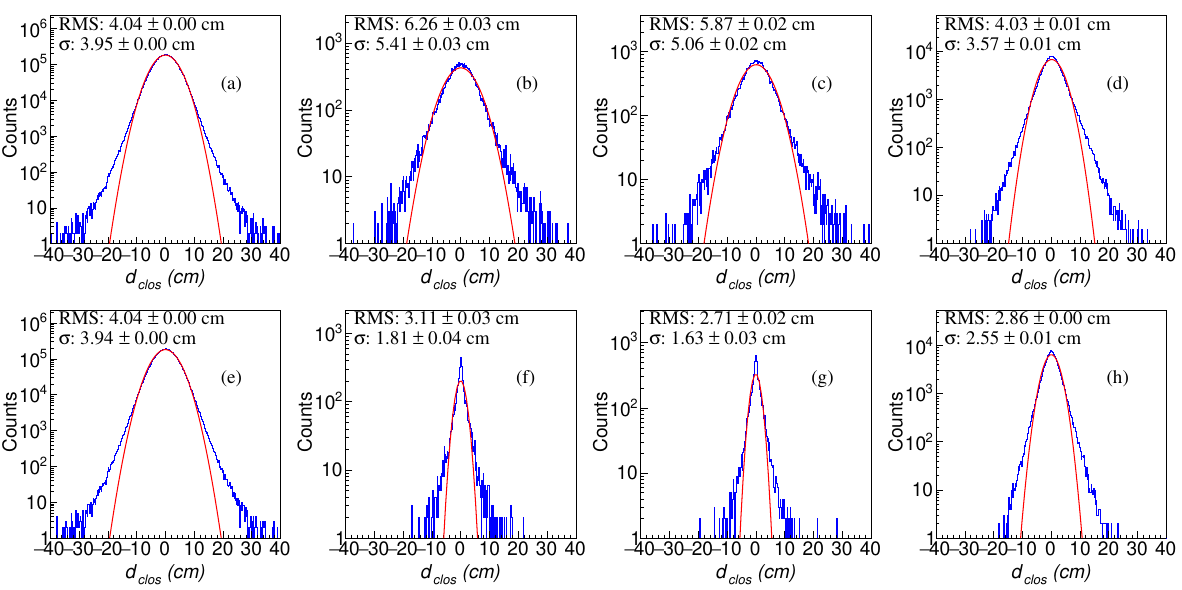}
  	\caption{Difference between extrapolated and expected muon position in each wall in the presence of a magnetic field; (a) Top wall, (b) Left wall, (c) Right wall and (d) Back wall.  Top rows are for all superclusters and bottom rows are for the minimum difference.}
  	\label{supclpos_mag}

  \end{figure}

  \begin{figure}[h]
  	\centering
  	\includegraphics[width=1.\textwidth]{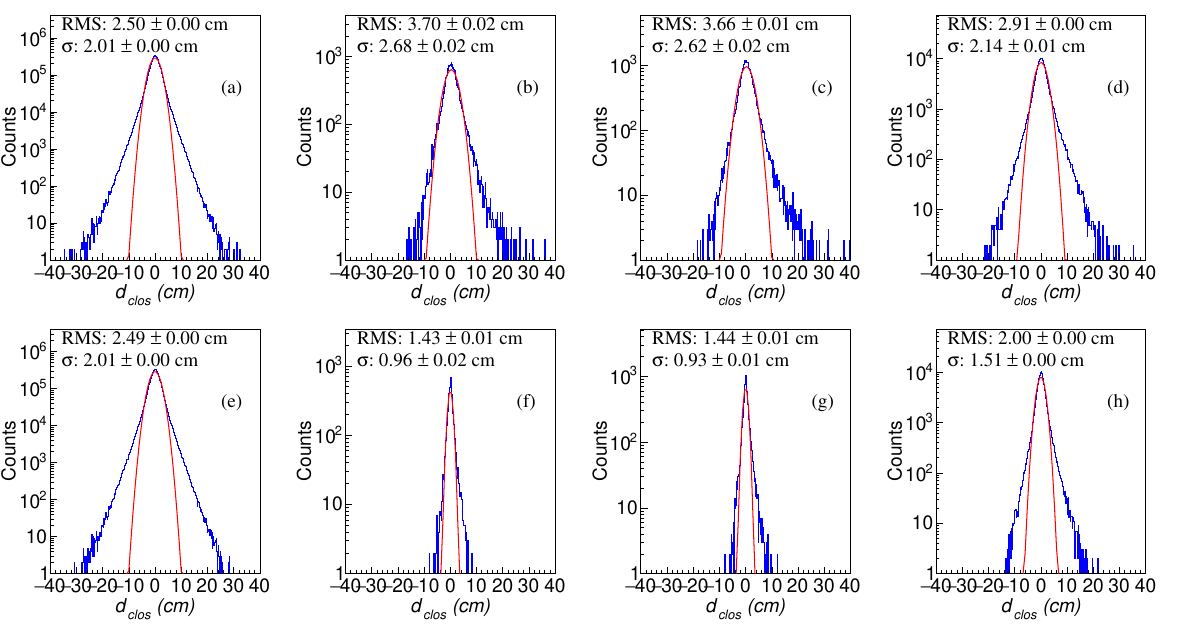}
  	
  	\caption{Difference between extrapolated and expected muon position in each wall in the absence of a magnetic field; (a,e) Top wall, (b,f) Left wall, (c,g) Right wall and (d,h) Back wall.}
  	\label{supclpos_nonmag}
  	
  \end{figure}

\begin{figure}
    \begin{center}
  	\includegraphics[width=0.5\linewidth]{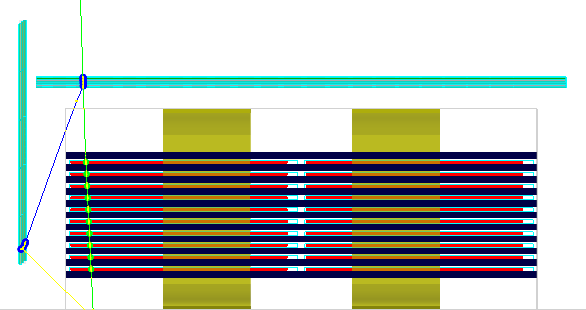}
    \end{center}
  	\caption{Valid Supercluster in multiple walls due to secondaries produced by muon interaction.}
  	\label{tailevents}
  \end{figure}

   Figures \ref{supclpos_mag} and \ref{supclpos_nonmag} show the closest distance in each wall with and without a magnetic field respectively.  The closest distance is considered positive if the extrapolated point in the plane is greater than the supercluster position in the miniICAL coordinate system as shown in figure \ref{miniicalcmvd}.
    In the Figures \ref{supclpos_mag} and \ref{supclpos_nonmag},  top row (figures: a-d), the closest distance is stored for all the four walls having a valid supercluster while in the bottom row (figures: e-h), it is stored only when that particular wall is nearest. The poor resolution of the two side walls (Left and Right) is attributed to their greater distance from the miniICAL detector and larger zenith angle of muon trajectory, where uncertainties are larger due to larger multiple scattering in irons as well as larger extrapolated lengths. Events in the tail part ( Figure \ref{supclpos_mag} and \ref{supclpos_nonmag},  top row (a-d)) are a result of hits generated from secondaries 
    as the muon interacts with detector/surrounding materials as shown in Figure \ref{tailevents}, particularly noticeable in the back-side wall since it is closest to the miniICAL detector.
    The skewed appearance of all the side walls is due to edge effects, where the supercluster is consistently contained within the boundary near the edges of the detector. 

\begin{figure}[h]
	\centering
	\includegraphics[width=1\textwidth]{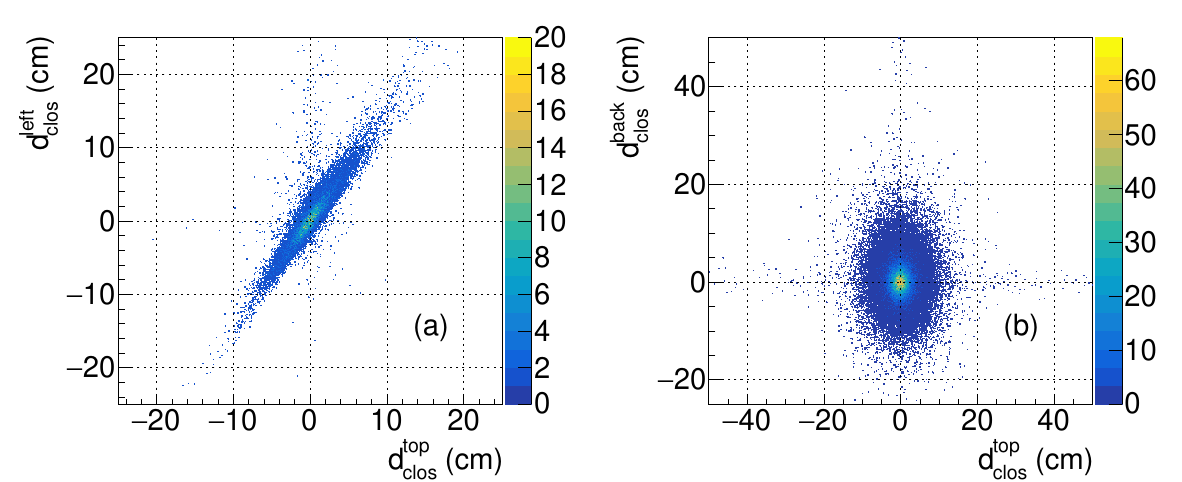}
	
	
        \caption{Closest distances between the reconstructed track and the supercluster line: (a) Superclusters reconstructed in both top and left walls, and (b) Superclusters in top and back walls. Events in the tail part originate from hits attributed to secondaries produced by muon interaction with detector material.}
        
	\label{multiwallhits}

\end{figure}

	However, when we look for the least closest distance among all walls, the left and right-side walls have better resolution. In the cases where these two side walls are nearest, the majority of events tend to have hits in the top wall due to the influence of trigger criteria and spatial arrangement of the walls. Given that the EPS strips in both the top and left/right walls are parallel, the closest distance measurements are correlated as shown in figure \ref{multiwallhits} (a). This correlation contributes to the observed improved resolution in the side walls when these walls were nearest. Conversely, for the back and top walls, where the EPS strips are perpendicular to each other, the uncertainties in the top wall do not significantly affect the uncertainties in the back wall. Due to the lack of correlation between the closest distances in the top and back walls as shown in figure \ref{multiwallhits} (b), there is no significant change in distributions when the back wall is the nearest one.
 
\begin{figure}[h]
	\begin{subfigure}{0.5\textwidth}
\includegraphics[width=1\textwidth]{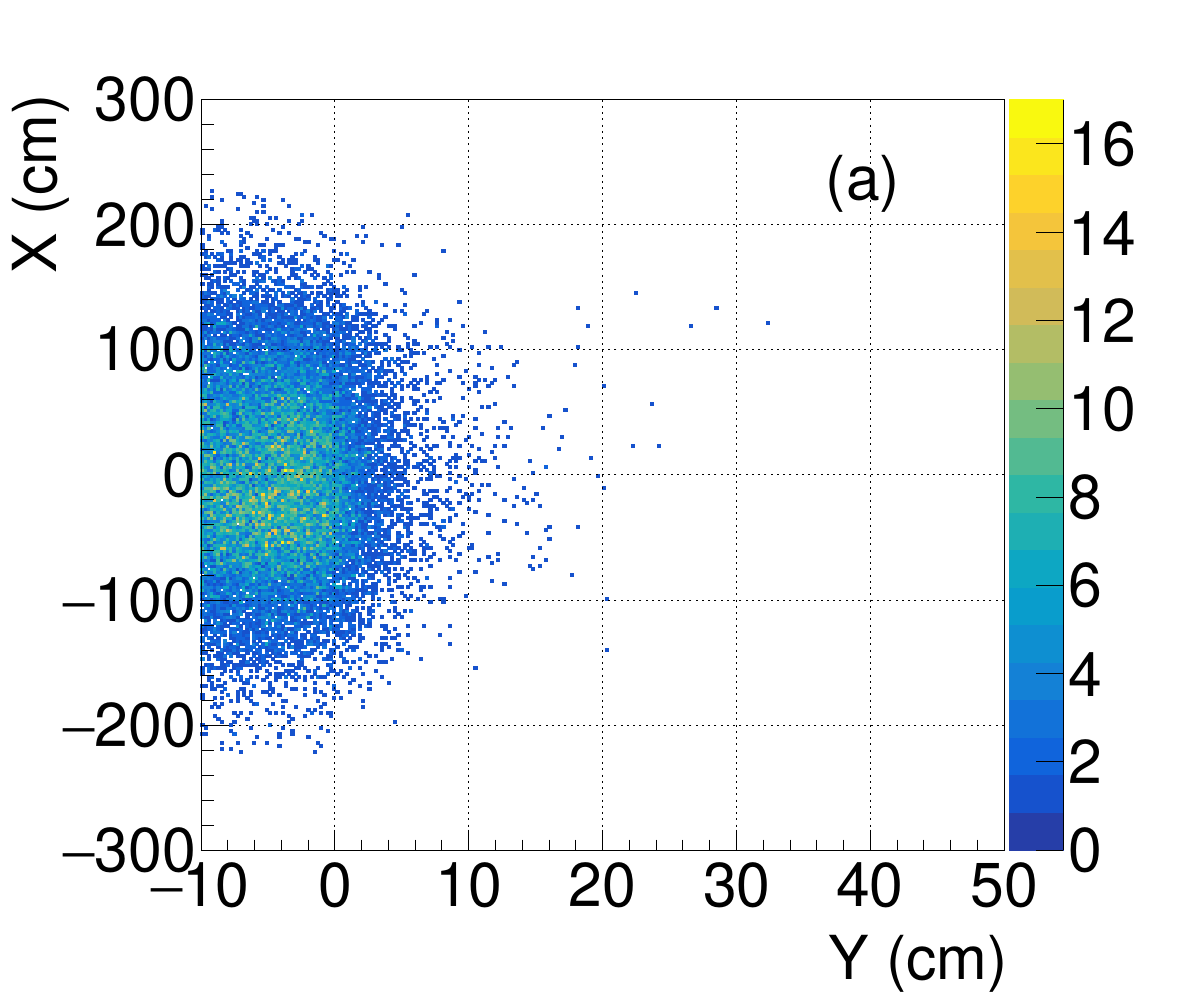}

	\end{subfigure}
\quad
	\begin{subfigure}{0.5\textwidth}
\includegraphics[width=1\textwidth]{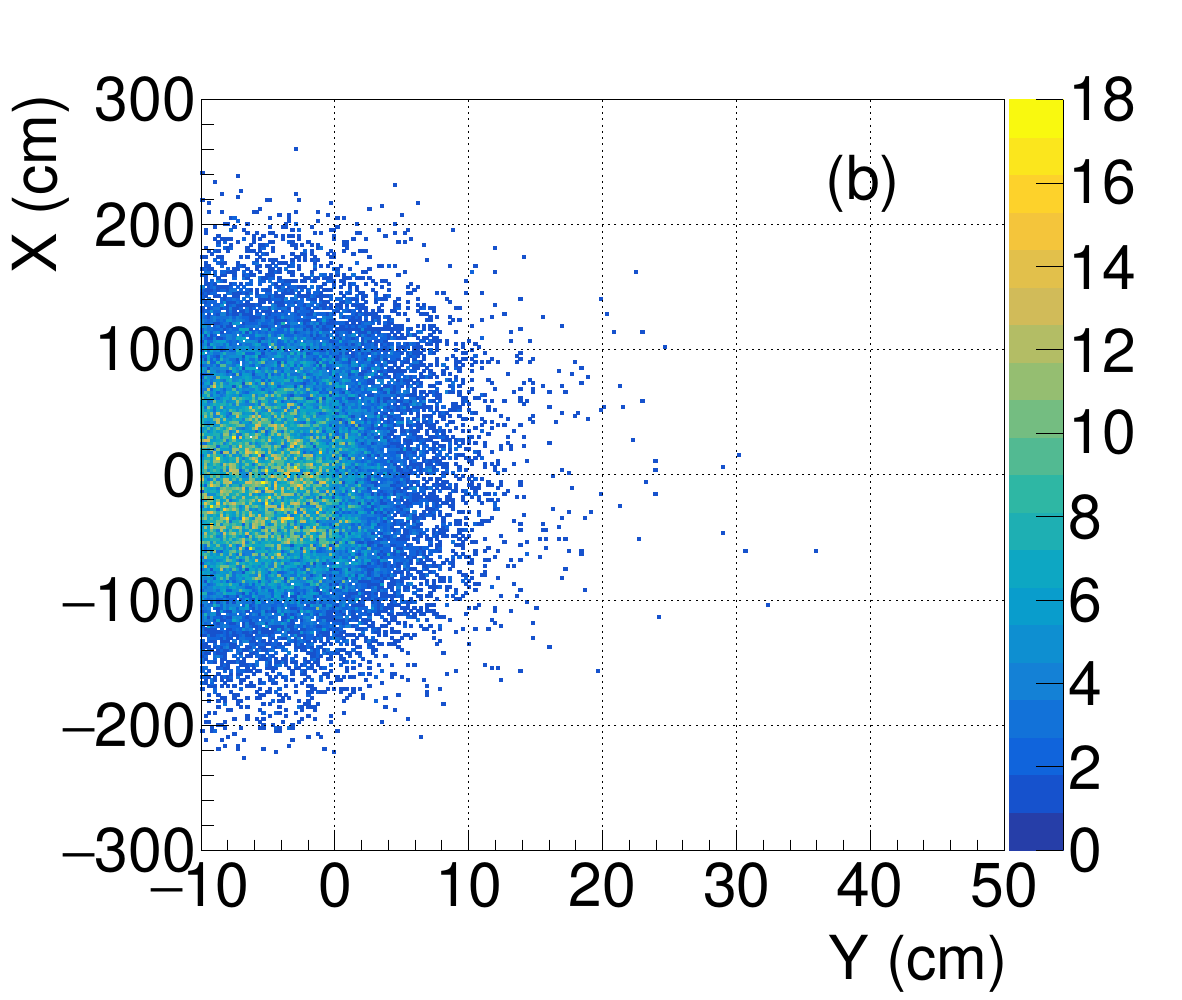}
		\end{subfigure}
\caption{Extrapolated muon position on the topmost layer when the muon trajectory misses the top wall: (a) without magnetic field, (b) with magnetic field.}
\label{fig:extrapol_nohits}  
\end{figure}

	The miniICAL RPC detectors sometimes require troubleshooting and repair. To allow access to the RPCs, the CMVD design excludes the front side veto wall. As a result, in the absence of the front wall, it is expected that there will be events where a muon did not pass through EPS strips, though there was a trigger generated in the miniICAL system.
        During the data-taking periods, the muon hits in the veto system will be identified by extrapolating the reconstructed muon trajectory toward the veto walls. However, it was found through simulation that there will be instances where the extrapolated position of the muon is inside the CMVD, even though the muon did not pass through that layer.
This issue was investigated and it was determined that in order to exclude such false detections, events with extrapolated positions within $30\,cm$ from the front-side of top veto wall must be disregarded. This reduces the acceptance of the cosmic muon veto detector but removes any ambiguity of mis-reconstructions.
This larger area is primarily due to factors such as the uncertainty in the reconstruction, the effect of multiple scattering within the iron layers, and the longer extrapolation distance for tracks with large angles. Figure  \ref{fig:extrapol_nohits} shows the extrapolated position on the veto walls for events originating from the front side of the detector, i.e., without any interaction in veto walls.

%

\section{Veto Criteria and Efficiency}
\label{chap_performance}  
Most of the events triggered in the miniICAL detector result in hits in the top wall. The two side walls (left and right) have fewer hits because of the specific trigger criteria mentioned in Section \ref{event_generation}.
The exact position of the muon in the CMVD wall is not precisely determined from the extrapolated position due to uncertainty in the reconstruction and multiple scattering with detector material.
The efficiency of CMVD is defined as the fraction of events with valid superclusters near the extrapolated muon positions in the CMVD detector. The muon track in the RPC stack with good fit quality is considered here to reduce the uncertainties in the extrapolated position in the CMVD. A supercluster as defined above must have at least two layers of nearby EPS detectors and 2 or more SiPMs associated with it must have a signal above 2.5 photo-electron. 
The "good fit quality" of a track is quantified by its $\chi^{2}$ and the total number of hits (nhits) in the RPC stack.

\begin{figure}[h]
	\begin{subfigure}{0.5\textwidth}
	\includegraphics[width=1.1\textwidth]{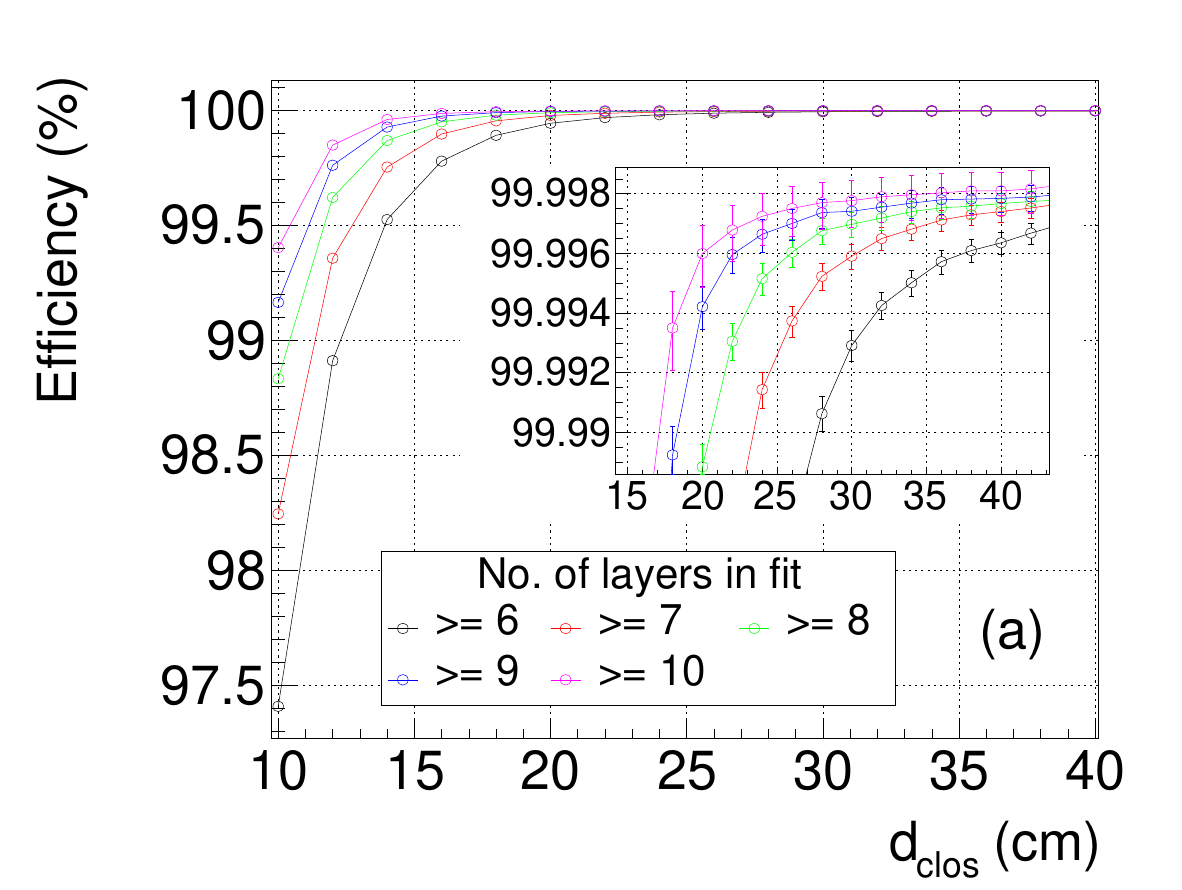}
	\end{subfigure}
\quad
	\begin{subfigure}{0.5\textwidth}
	\includegraphics[width=1.1\textwidth]{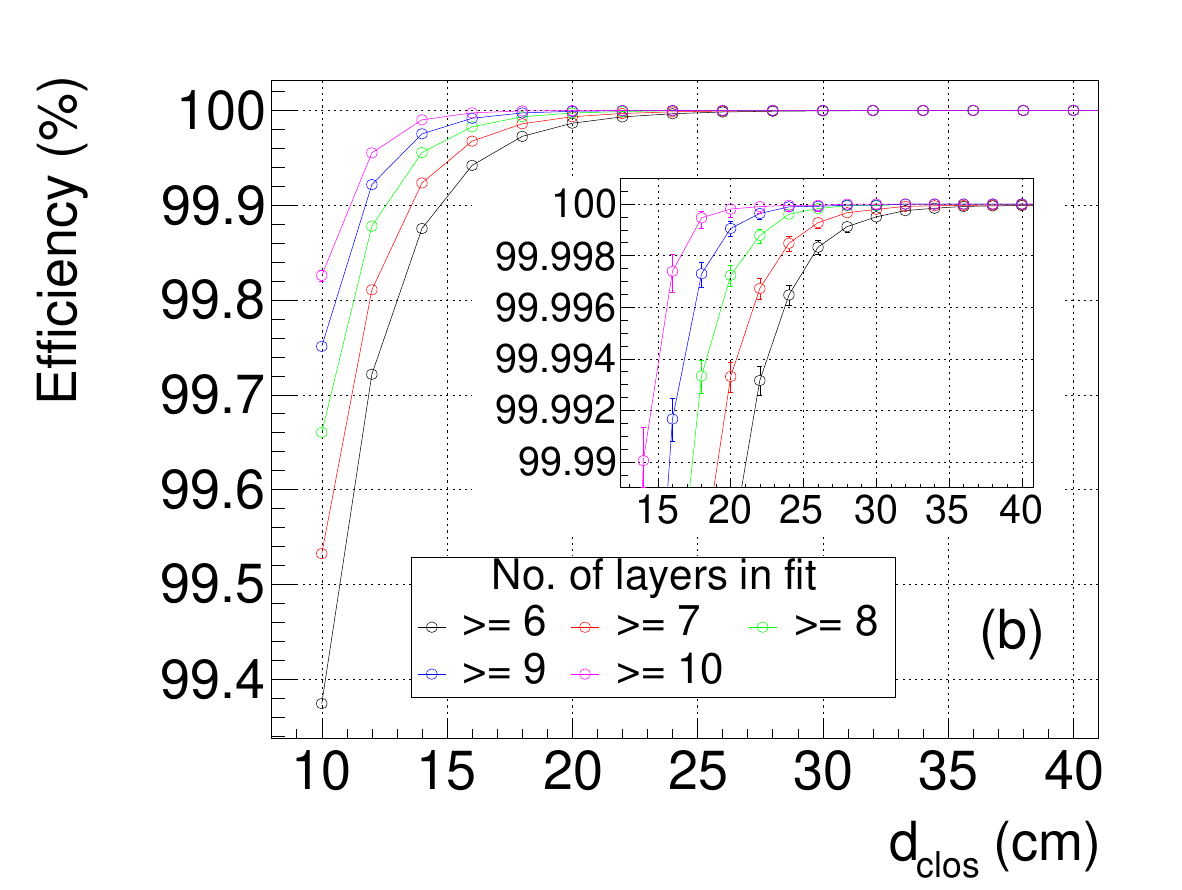}
		\end{subfigure}
\caption{Efficiency versus closest distance for different number of layers in fitted track: (a) with magnetic field, (b) without magnetic field. The SiPM threshold to form a hit was kept at 2.5 p.e. equivalent.}
\label{effi_vs_diff}
\end{figure}

\begin{figure}[h]
\centering
\begin{minipage}{0.45\textwidth}
    \centering
    \includegraphics[width=\textwidth]{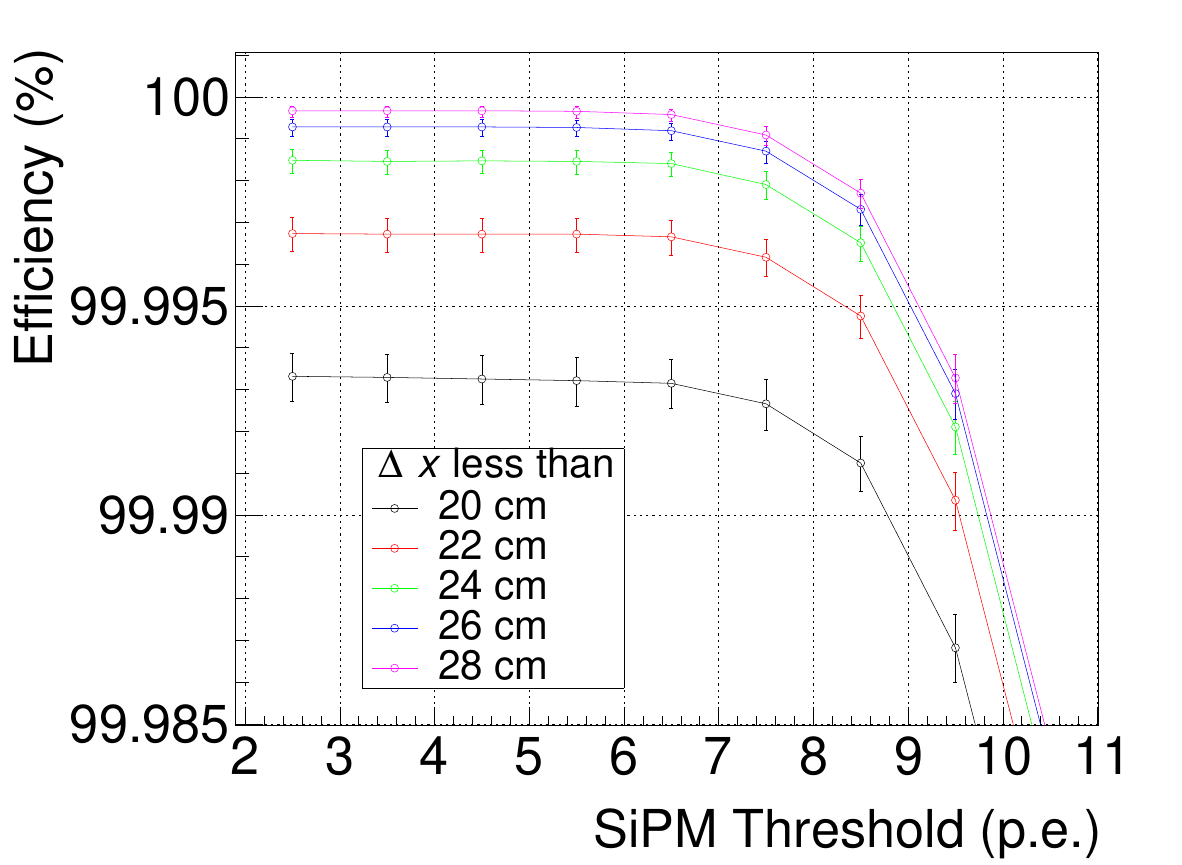}
    \caption{Efficiency versus different SiPM thresholds for different search windows for well-fitted tracks ($\chi^{2}/ndof<2$ and number of layers $\geq$ 7).}
    \label{effi_vs_thres}
\end{minipage}\hfill
\begin{minipage}{0.45\textwidth}
    \centering
    \includegraphics[width=\textwidth]{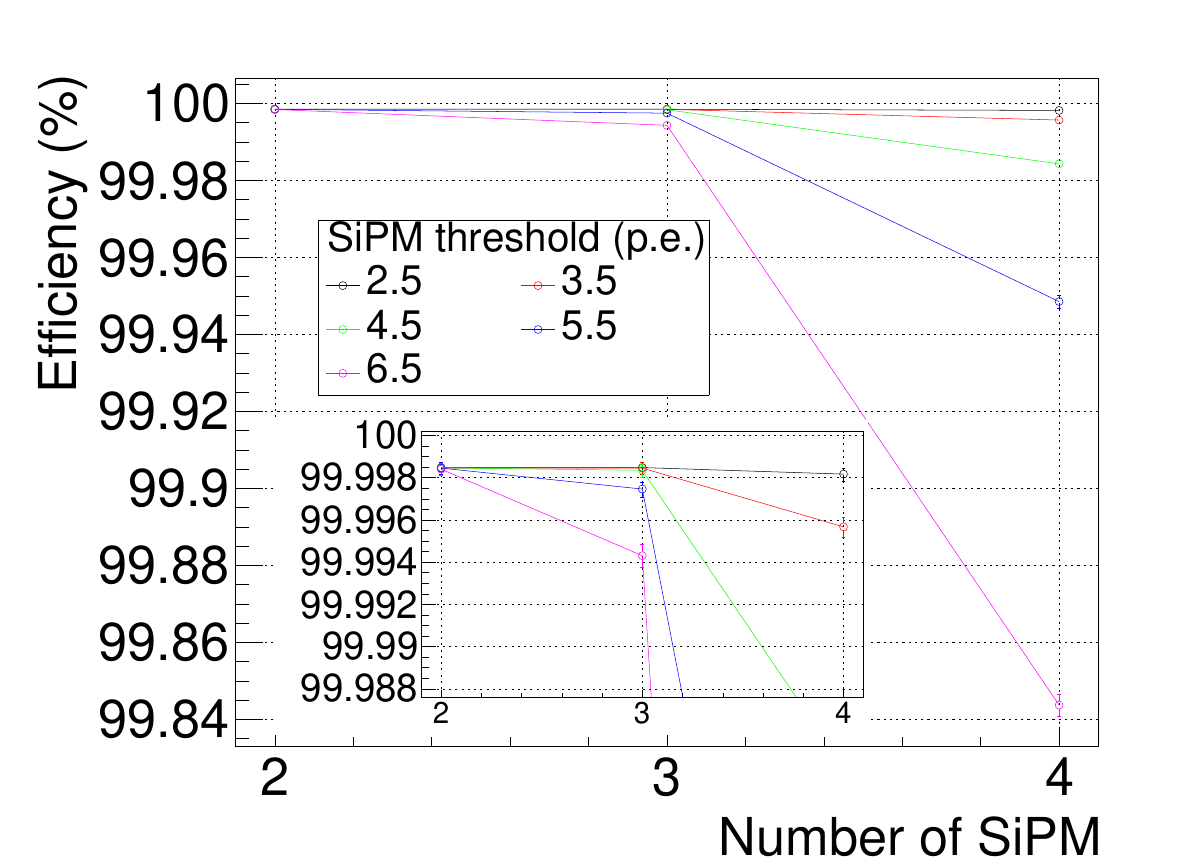}
    \caption{Efficiency versus different number of SiPMs used to form a hit for different SiPM thresholds for well-fitted tracks ($\chi^{2}/ndof<2$ and number of layers $\geq$ 7) and search window ($\Delta x$) within $30\,cm$.}
    \label{effi_vs_nsipm}
\end{minipage}
\end{figure}

Figure \ref{effi_vs_diff} shows the efficiency as a function of the distance between the extrapolated track position and supercluster position (search window) in the veto wall with and without the presence of a magnetic field.
Since the extrapolated position depends on the uncertainty in the reconstruction of the muon, a higher number of RPC layers in the muon track reconstruction leads to better efficiency. 
To achieve an efficiency of more than 99.99\,\%, we must increase our search area to at least $30\,cm$ ($25\,cm$ without a magnetic field) from the extrapolated position for all the selected reconstructed muon tracks in the RPC stack.
Additionally, within the current simulation, which includes approximately 8.5 million selected muons, we did not observe any false positive events.
Figure \ref{effi_vs_thres} illustrates the variation in efficiency for different SiPM threshold values used in the simulation for hit formation, considering different values of the search criteria. The veto efficiency is found to remain relatively constant for thresholds up to 6.5 photo-electrons. Figure \ref{effi_vs_nsipm} shows the variation of efficiency for different numbers of SiPMs used in hit formation considering various thresholds. All four SiPMs can be utilized for hit formation up to 3.5 photo-electron threshold, while any three SiPMs can be used for thresholds up to 6.5 photo-electrons. However, in the realistic experimental scenario, we do not expect all SiPM will performs like an ideal scenario, thus the selection criteria for the CMVD strips signal remains the same as it was studied earlier \cite{Jangra:2021key}.

\section{Conclusion}
 In this paper, the construction and operation of a Cosmic Muon Veto Detector (CMVD) is described. A dedicated software tool has been developed for the CMVD, tuned using inputs from experimental data, and integrated with the existing mini-ICAL apparatus, consisting of RPC detectors. The muon tracks reconstructed in the RPCs are used to estimate CMVD efficiency, and consequently establish effective muon veto criteria. This criteria takes into account the noise level in the SiPM detector and the gap between the EPS strips.
 There should be signals above 2.5 p.e. equivalent in at least two SiPMs in each of these strips and at least 2 layers must have a valid signal. This veto criteria have been evaluated for well-fitted tracks ($\chi^{2}/\text{ndf}<2$ and $\text{nHits}>6$) within the RPC stack. Initially, tracks with extrapolated points within $30\,cm$ from the front side must be rejected to account for the absence of the front wall. Subsequently, if a reconstructed track's extrapolated position lies within the detector wall and a hit is detected within $30\,cm$ from the extrapolated position, it is considered a muon hit. For tracks outside the detector region, a muon hit is considered if the distance of closest approach between the track and the detector wall edge is within $30\,cm$. Furthermore, the veto efficiency remains nearly unchanged up to 6.5 p.e. SiPM threshold. Additionally, more than 2 SiPMs in an EPS will register a signal above this threshold, which rejects most of the uncorrelated SiPM noise. The fake rate in efficiency measurement is negligible. Therefore, if the EPS strips are light leak tight and maintained in the same dark environment as the test setup, the veto system can achieve the desired efficiency.

\section{Acknowledgments}
We sincerely thank all the mini-ICAL group members at IICHEP, Madurai. We would also like to
thank other members of the INO collaboration for their valuable inputs.


\bibliographystyle{ieeetr} 
\bibliography{refrences.bib}

\begin{thebibliography}{10}

\bibitem{Majumder:2018xel}
G.~Majumder and S.~Mondal, ``{Design, construction and performance of
  magnetised mini-ICAL detector module},'' {\em PoS}, vol.~ICHEP2018, p.~360,
  2019.

\bibitem{Panchal:2017aub}
N.~Panchal, S.~Mohanraj, A.~Kumar, T.~Dey, G.~Majumder, R.~Shinde, P.~Verma,
  B.~Satyanarayana, and V.~M. Datar, ``{A compact cosmic muon veto detector and
  possible use with the Iron Calorimeter detector for neutrinos},'' {\em
  JINST}, vol.~12, no.~11, p.~T11002, 2017.

\bibitem{Saraf:2023sjs}
M.~Saraf {\em et~al.}, ``{Design, fabrication and large scale qualification of
  cosmic muon veto scintillator detectors},'' {\em JINST}, vol.~18, no.~05,
  p.~P05003, 2023.

\bibitem{Jangra:2021key}
M.~Jangra {\em et~al.}, ``{Characterization of Silicon-Photomultipliers for a
  Cosmic Muon Veto detector},'' {\em JINST}, vol.~16, no.~11, p.~P11029, 2021.

\bibitem{Pla-Dalmau:2000puk}
A.~Pla-Dalmau, A.~D. Bross, and K.~L. Mellott, ``{Low-cost extruded plastic
  scintillator},'' {\em Nucl. Instrum. Meth. A}, vol.~466, pp.~482--491, 2001.

\bibitem{Alekseev:2021vbe}
I.~Alekseev, M.~Danilov, V.~Rusinov, E.~Samigullin, D.~Svirida, and
  E.~Tarkovsy, ``{The performance of a new Kuraray wavelength shifting fiber
  YS-2},'' {\em JINST}, vol.~17, no.~01, p.~P01031, 2022.

\bibitem{sipm}
``{HAMAMATSU MPPC (Multi-Pixel Photon Counter) datasheet}.''
\newblock
  \url{https://www.hamamatsu.com/eu/en/product/type/S13360-2050VE/index.html}.

\bibitem{Agostinelli:2002hh}
S.~Agostinelli {\em et~al.}, ``{GEANT4: A simulation toolkit},'' {\em Nucl.
  Instrum. Meth.}, vol.~A506, pp.~250--303, 2003.

\bibitem{Heck:1998vt}
D.~Heck, J.~Knapp, J.~N. Capdevielle, G.~Schatz, and T.~Thouw, ``{CORSIKA: A
  Monte Carlo code to simulate extensive air showers},'' 2 1998.

\bibitem{John:2022fuy}
J.~M. John, S.~Pethuraj, G.~Majumder, N.~K. Mondal, K.~C. Ravindran, V.~M.
  Datar, and B.~Satyanarayana, ``{Improving time and position resolutions of
  RPC detectors using time over threshold information},'' {\em JINST}, vol.~17,
  no.~04, p.~P04020, 2022.

\bibitem{Birks:1951boa}
J.~B. Birks, ``{Scintillations from Organic Crystals: Specific Fluorescence and
  Relative Response to Different Radiations},'' {\em Proc. Phys. Soc. A},
  vol.~64, pp.~874--877, 1951.

\bibitem{drs4}
``{DRS4, DRS4 Evaluation Board User's Manual}.''
\newblock
  \url{https://www.psi.ch/sites/default/files/2020-02/manual_rev51.pdf}.

\bibitem{Bhattacharya:2014tha}
K.~Bhattacharya, A.~K. Pal, G.~Majumder, and N.~K. Mondal, ``{Error propagation
  of the track model and track fitting strategy for the Iron CALorimeter
  detector in India-based neutrino observatory},'' {\em Comput. Phys. Commun.},
  vol.~185, pp.~3259--3268, 2014.

\bibitem{Pal:2014tre}
S.~Pal, {\em {Development of the INO-ICAL detector and its physics potential}}.
\newblock PhD thesis, HBNI, Mumbai, 2014.

\end{thebibliography}

\end{document}